\documentclass[12pt]{article}
\usepackage{epsfig}
\usepackage{array}
\usepackage{float}
\usepackage{amstext}
\usepackage{rotating}
\usepackage{a4}
\usepackage{a4wide}
\usepackage{cite}

\newcommand{\be}{\begin{eqnarray}}
\newcommand{\ee}{\end{eqnarray}}

\def\nn{\nonumber}

\begin{document}

\title{
\Large 
Spontaneous R-Parity Violation, $A_4$ Flavor Symmetry and Tribimaximal Mixing}
\author{
Manimala Mitra\thanks{email: \tt mmitra@hri.res.in}
\\\\
{\normalsize \it Harish--Chandra Research Institute,}\\
{\normalsize \it Chhatnag Road, Jhunsi, 211019 Allahabad, India }\\ \\ 
}
\date{ \today}
\maketitle
\vspace{-0.8cm}
\begin{abstract}
\noindent

We explore the possibility of spontaneous R parity violation in the context of $A_4$ flavor symmetry. 
Our model contains $SU(3)_c \times SU(2)_L \times U(1)_Y$ singlet matter chiral superfields which are arranged as 
triplet of $A_4$ and as well as few additional Higgs chiral superfields which are singlet under 
MSSM gauge group and belong to triplet and singlet representation under the $A_4$ flavor symmetry. 
R parity is   broken spontaneously by the vacuum expectation values of the different 
sneutrino fields and hence we have neutrino-neutralino as well as neutrino-MSSM gauge  singlet 
higgsino mixings in our model, in addition to the standard model neutrino- gauge singlet neutrino, gaugino-higgsino and higgsino-higgsino mixings.
Because  all of these mixings we have an extended neutral fermion mass matrix.
We explore the low energy neutrino mass matrix for our model and point out  that with some specific constraints between 
the sneutrino vacuum expectation values as well as the MSSM gauge singlet  Higgs vacuum expectation values,  the low energy neutrino mass matrix will lead to a tribimaximal mixing matrix. We also analyze the potential minimization for our model and show that one can realize a higher vacuum expectation value of  the $SU(3)_c \times SU(2)_L \times U(1)_Y$  singlet sneutrino fields even when the other sneutrino vacuum expectation values are  extremely small or even zero.  

\end{abstract}

\newpage

\section{Introduction}

Various  experimental evidences on neutrino mass and mixing have opened up a window to physics beyond the standard model of particle physics. Experiments like SNO, KamLAND, K2K and MINOS \cite{solar, kl, k2k, minos} provide information on the two mass square differences $\Delta m_{21}^2$ and $\Delta m_{31}^2$ and on the two mixing angles $\theta_{12}$ and $\theta_{23}$. The third mixing angle $\theta_{13}$ is yet not determined, but certainly is known to be small \cite{chooz}. The current $3 \sigma$  allowed intervals  of the oscillation parameters are  given as \cite{limits}

\be
 7.1 \times 10^{-5} \rm{eV^2}<\Delta m_{21}^2 < 8.3\times 10^{-5} \rm{eV^2},\hspace*{0.1cm} 2.0 \times 10^{-3} \rm{eV^2}<\Delta m_{31}^2 < 2.8\times 10^{-3} \rm{eV^2}\ee 
\be
   0.26< \sin^2 \theta_{12}<0.42,\hspace*{0.1cm}    0.34 <\sin^2 \theta_{23}<0.67 , \hspace*{0.1cm}  \sin^2 \theta_{13}<0.05 \ee 

To explain the above mentioned very precise data of   neutrino mass and mixing and  without bringing any additional fine tuning problem into the theory, one has to has look for beyond standard model physics. In beyond standard model physics very small Majorana 
neutrino masses can be generated by the 
dimension 5 operator $\frac{1}{\Lambda}LLHH$ \cite{dim5}, 
where the masses are suppressed naturally by the scale of new 
physics $\Lambda$. Note that this term breaks lepton number 
 which is mandatory for the generation of 
Majorana masses. The seesaw \cite{seesaw} in its simplest version could be Type I seesaw \cite{seesaw}, Type II \cite{type2} or Type III\cite{type3,type3all,type3others,type3us} depending on the heavy particles which would be integrated out and generate the above mentioned dimension 5 operator are standard model singlet, standard model triplet  with hypercharge $Y=2$ and standard model  triplet with hypercharge $Y=0$ respectively.   
Observed neutrino mixing can be obtained very naturally by imposing
flavor symmetry. Among the numerous viable flavor symmetry models 
\cite{flvgen}, the models based on the group $A_4$ are  the most popular ones
\cite{flv, flvalta, flvnum,Feruglio:2009iu }.

Among all the different models of beyond standard model physics supersymmetry is probably the most attractive one, for its capability to solve the Higgs mass hierarchy problem very naturally. The most general superpotential which respects the standard model gauge group also  allows lepton number and baryon number violating bilinear $\epsilon \hat{L}\hat{H}_u$  and trilinear $\lambda, \lambda^{\prime}, \lambda^{\prime \prime}$  terms \cite{Rparity, Rparityold, Rparityold2}. The R-Parity is a discrete symmetry defined as $R_p=(-1)^{3(B-L)+2S}$ and for all matter chiral superfields it is $-1$ while for Higgs chiral superfields it is $+1$. Defined  in this way  R-parity conservation forbids all the baryon and lepton number violating terms in the superpotential.  However  minimal supersymmetric standard model with   R parity violation \cite{gauginoseesaw,spontaneous,Roy-mu,Roy:1996bua,susybneu, extrarightnu,numssm} opens up the possibility of neutrino mass generation. If one sticks to the basic MSSM gauge group and to the MSSM particle contents and breaks R parity spontaneously, then eventually one will encounter with the problem of Majoron \cite{spontaneous}, which could be evaded if one extends the MSSM particle contents and/or extends the gauge group suitably.  

The neutrino masses could be generated from the bilinear  as well as from the trilinear R-Parity violating terms of the superpotential. While the bilinear R-Parity violating  $\hat{L}\hat{H}_u$ term generates  the neutrino mass via neutrino-higgsino mixing \cite{Roy-mu,Roy:1996bua}, neutrino masses could also be generated from the  lepton number violating trilinear terms via loop effect \cite{oneloop, susybneu1, twoloop}. However severe bounds \cite{bfact} on the lepton and baryon number violating  $\lambda$, $\lambda^{\prime}$ and $\lambda^{\prime \prime}$ terms of the superpotential come from the non-observation of proton decay which essentially constraint the simultaneous presence of  lepton number  and baryon number violation   in the superpotential.

The spontaneous R parity violation provides a natural explanation for the absence of baryon number violating $\lambda^{\prime \prime}$ term in the superpotential, as long as one sticks to the renormalizable field theory. In the scheme of spontaneous R parity violation  R parity is a symmetry of the theory and once different sneutrino fields get vacuum expectation values, R parity is being violated spontaneously. In addition to the field contents of MSSM  if one has a MSSM gauge singlet/triplet  sneutrino state, then the bilinear R parity violating term involving leptons and Higgs would be generated once this gauge singlet/triplet sneutrino state gets vacuum expectation value. It is possible to realize  the other terms $\lambda$ and $\lambda^{\prime}$ from the R parity conserving MSSM superpotential only after redefinition of the basis. 

There have been several attempts to realize  R-parity violation spontaneously. In the context of a  GUT theory one can relate the spontaneous R-parity violation with the gauge symmetry breaking \cite{pavel} and it is possible to  realize some of the sneutrino vacuum expectation values to be in the TEV scale. In these kind of models the low energy neutrino masses  would be generated via double seesaw mechanism. However if one sticks to the basic  MSSM gauge group and explore the possibility of spontaneous R-parity violation without invoking any problem of Majoron, one has to extend the particle content of the model. One can introduce the standard model singlet/triplet matter chiral  superfield into the theory. In this kind of model the low energy neutrino mass would be generated via the Type-I/Type-III seesaw and it is possible to get proper mass splitting between the low energy neutrino masses and the correct mixing even with one generation of heavy neutrino matter chiral superfield \cite{Choubey:2009wp}. The different sneutrino vacuum expectation values share a proportionality relation in this kind of model. 

In this work we explore the possibility of spontaneous R-parity violation in the context of  $A_4$ flavor symmetry while we stick to the MSSM gauge group. The best fit value of the neutrino oscillation parameters points towards a tribimaximal neutrino mixing matrix,  which is possible to achieve very naturally for the R parity conserving scenario if one imposes a flavor symmetry such as $A_4$ in the theory \cite{flvalta}. In our model we have few additional standard model singlet Higgs superfields $\hat{\phi}_T$, $\hat{\phi}_S$ and $\hat{\xi}$ along with the  standard model singlet matter chiral superfields $\hat{N}$. In addition to $A_4$  we have also implemented one  $Z_3$ symmetry in our model. The different Higgs chiral superfields and matter chiral superfields belong to the triplet as well as singlet representation under the flavor symmetry group $A_4$ and have suitable $Z_3$ charges. The symmetry group $A_4$ would be broken by the vacuum expectation value of the $A_4$ triplet fields while R-parity would be broken spontaneously by the vacuum expectation values  of the different sneutrino $\tilde{\nu}$ and $\tilde{N}$ fields. The neutral fermion mass matrix in our model is enlarged compared to that of minimal supersymmetric standard model and  in addition to the conventional Dirac mixing between the standard model neutrino $\nu$ and the gauge singlet neutrino $N$,  R-parity violation brings  mixing between the different neutrino and neutralino states. In our model because of R parity violation we also have other mixings between the different neutrino states  and the MSSM gauge singlet higgsino fields along with the different gaugino-higgsino and higgsino-higgsino mixings. With all these mixings we explore the possibility of generating tribimaximal mixing in our model. We  point out   that although the neutral fermion mass matrix is enlarged and has many parameters, however the low energy neutrino mass matrix would still lead to a tribimaximal mixing matrix, provided  the vacuum expectation values of the different sneutrino and gauge singlet Higgs satisfy some specific constraints. We also show that in  this model it is possible to get a higher vacuum expectation value of the gauge singlet sneutrino fields  $ \tilde{N} $  even if the other sneutrino vacuum expectation values $\langle \tilde{\nu} \rangle$ are  extremely small and even  $\langle \tilde{\nu} \rangle \to 0$. 

The paper is organized as follows. In section 2 we describe the model and in section 3 we discuss the neutrino phenomenology where we explore the possibility of getting tribimaximal mixing for our model. In section 4 we discuss the symmetry breaking  part where we show that even in  the $\langle \tilde{\nu} \rangle \to 0$ limit one expects  a higher value of the sneutrino vacuum expectation value $\langle \tilde{N}\rangle $. In section 4 we present our conclusion. Discussion on the soft supersymmetry breaking Lagrangian for this model has been 
presented in the Appendix.

\section{Model}

In this section we present our model. We stick to the $SU(3)_c \times SU(2)_L \times U(1)_Y$ gauge group of MSSM and  impose $A_4$ flavor symmetry in our model. In addition to the  $A_4$ triplet and standard model singlet Majorana superfields $\hat{N}$, our model contains  one standard model and $A_4$ singlet superfield  $\hat{\xi}$ and two $A_4$  triplets and standard model singlet superfields  $\hat{\phi}_T$ and  $\hat{\phi}_S$. The particle contents of our model has  explicitly been given in Table. (\ref{tab:parc}). The details about the  discrete symmetry group $A_4$ could be found in  \cite{flvalta,Feruglio:2009iu }. With these superfield contents the superpotential of our  model is,
\be
W &=&  y_1\hat{e}^c(\hat{\phi}_TL)+ y_2\hat{\mu}^c(\hat{\phi}_TL)^{\prime \prime}
+ y_3\hat{\tau}^c(\hat{\phi}_TL)^{\prime }+ y_{\nu}\hat{N}\hat{L}\hat{H}_u  \nn \\ && + x_A\hat{\xi}(\hat{N}\hat{N})+ x_B(\hat{\phi}_S \hat{N}\hat{N})+ \alpha \hat{\xi}\hat{\xi}\hat{\xi} + \beta^{\prime} \hat{\phi}_S\hat{\phi}_S \hat{\phi}_S + \beta \hat{\phi}_S\hat{\phi}_S \hat{\xi} \nn \\ && +\gamma \hat{\phi}_T\hat{\phi}_T  + \gamma^{\prime} \hat{\phi}_T \hat{\phi}_T \hat{\phi}_T + \mu \hat{H}_u\hat{H}_d .
\label{eq:wsup}
\ee 
Note that apart from  $A_4$, one more  discrete $Z_3$ symmetry has been implemented  in our model so that $\hat{\phi}_T$ does not contribute to the standard model neutrino mass generation, $\hat{\xi}$  and $\hat{\phi}_S$ do not  contribute to the charged lepton masses,  although $\hat{\xi}$  and $\hat{\phi}_S$ contribute to the neutrino sector significantly.  We represent the superfields  $\hat{\xi}$ and $\hat{N}$  as follows,
\be
\hat{\xi}=\xi+\sqrt{2} \theta \tilde{\xi}+ \theta \theta F_{\xi},
\label{eq:chi}
\ee
\be
\hat{N}_i=\tilde{N}_i+\sqrt{2} \theta N_i +\theta \theta F_{N_i}.
\label{eq:chin}
\ee
where $i$ represents the $A_4$ index and varies from $i=1,2,3$. The other superfields $\hat{\phi}_{T}$ and $\hat{\phi}_{S}$ will have the same structure as of $\hat{\xi}$ given in Eq. (\ref{eq:chi}) i.e, $\phi_{T_i,S_i}$ denote  the scalar partners and $\tilde{\phi}_{T_i,S_i}$ denote the fermionic partners and $F_{T_i,S_i}$ are the auxiliary components. 
\begin{table}[h]
\begin{center}
\begin{tabular}{|c|c|c|c|c|c|c|c|c|c|}
\hline
Field & $\hat{L}$  & $\hat{e}^c$  & $\hat{\mu}^c$  & $\hat{\tau}^c$  & $\hat{N}$ & $\hat{H}_{u,d}$ & $\hat{\xi}$ & $\hat{\phi}_T$ & $\hat{\phi}_S$ \cr
\hline
$A_4$ &   3 & 1 & $1^{\prime}$ & $1^{\prime \prime}$ & $3$ & 1 & 1 & $3$ & $3$ \cr
\hline
$Z_3$ & $\omega$ & $\omega^2$ & $\omega^2$ &  $\omega^2$ &  $\omega^2$ & 1  &  $\omega^2$ & 1 &  $\omega^2$ \cr
\hline
$R_p$ & -1 & -1 & -1 & -1 & -1 & +1 & +1 & +1 & +1 \cr
\hline
\end{tabular}
\caption{\label{tab:parc}
Field  transformation under $A_4$ and $Z_3$
}
\end{center}
\end{table}
Note that the  matter chiral superfields $\hat{N}$, $\hat{L}$, $\hat{e}^c$,  $\hat{\mu}^c$ and $\hat{\tau}^c$ are odd under R-parity while the two usual MSSM Higgs chiral superfields $\hat{H}_{u,d}$ as well as the other MSSM singlet Higgs chiral superfields $\hat{\phi}_T$, $\hat{\phi}_S$ and $\hat{\xi}$ are even under R parity. The superpotential given in Eq. (\ref{eq:wsup}) as well as the K\"ahler potential given in Eq. (\ref{eq:kah1}) and Eq. (\ref{eq:kah2}) conserve  R-parity,  however  R-parity will be broken spontaneously  by the vacuum expectation values of the different sneutrino fields $\tilde{N_i}$ and $\tilde{\nu}_i$. We would like to point out here that the $y_1\hat{e}^c(\hat{\phi}_TL)$ term in the superpotential Eq. (\ref{eq:wsup})  actually represents $\frac{y_1}{\Lambda} \hat{H}_d(\hat{\phi}_TL)\hat{e}^c$ and similarly for the other operators $y_2\hat{\mu}^c(\hat{\phi}_TL)^{\prime \prime}$ and $y_3\hat{\tau}^c(\hat{\phi}_TL)^{\prime }$. We would also like to stress here that in our model the lepton number is explicitly broken in the superpotential by few of the the trilinear  terms. Hence the sponateneous  R parity violation is not associated with any global $U(1)$ symmetry breaking. Therefore in our model sponateneous  R parity violation does not bring any problem of Majoron. 

In the R parity conserving scenario, the MSSM neutralinos  $\tilde{\lambda}^{0,3}$, $\tilde{H}^0_{u,d}$ as well as the higgsinos of the other MSSM gauge singlet Higgs chiral superfields i.e $\tilde{\phi}_T$, $\tilde{\phi}_S$ and $\tilde{\xi}$  decouple from the  neutrino sector $\nu$ and  $N$. The low energy neutrino mass matrix for R-parity conserving scenario would be,
\be
M_{\nu} \sim m_D^T\tilde{M}^{-1}m_D,
\ee
where 
\be
m_D=y_{\nu}v_2\pmatrix {1 & 0 & 0 \cr 0 & 0 & 1 \cr 0 & 1 & 0},
\ee
and 
\be
\tilde{M}=2\pmatrix{ a+\frac{2b_1}{3} & -\frac{b_3}{3} & -\frac{b_2}{3} \cr  -\frac{b_3}{3} & \frac{2b_2}{3} & a-\frac{b_1}{3} \cr -\frac{b_2}{3} & a-\frac{b_1}{3} & \frac{2b_3}{3}}.
\ee
$a$ and $b_i$ are $ a=x_As$  and   $b_i=x_Bu_i$ respectively where  $\langle \xi \rangle =s$ and $\langle \phi_{S_i}\rangle= u_i$. In the limit $u_1=u_2=u_3=u$, the mass matrix $\tilde{M}$ has this following form,
\be
\tilde{M}=2\pmatrix{ a+\frac{2b}{3} & -\frac{b}{3} & -\frac{b}{3} \cr  -\frac{b}{3} & \frac{2b}{3} & a-\frac{b}{3} \cr -\frac{b}{3} & a-\frac{b}{3} & \frac{2b}{3}},
\ee
where $b=x_Bu$ and the low energy neutrino mass matrix becomes,
\be
M_{\nu}=\frac{y_{\nu}^2v_2^2}{3a(b^2-a^2)}\pmatrix{b^2+2ab-3a^2 & b^2-ab & b^2-ab \cr b^2-ab & b^2+2ab & b^2-ab-3a^2 \cr  b^2-ab & b^2-ab-3a^2 & b^2+2ab}.
\ee
The  mixing matrix $U_{\nu}$ which diagonalizes the mass matrix $M_{\nu}$ and satisfies the diagonalizing relation  $U^TM_{\nu}U^*=D_k$ has this  tribimaximal form,
\be
U=\pmatrix{ \sqrt{\frac{2}{3}} & \sqrt{\frac{1}{3}} & 0 \cr  \sqrt{\frac{1}{6}} & \sqrt{\frac{1}{3}} & -\sqrt{\frac{1}{2}} \cr  \sqrt{\frac{1}{6}} &  \sqrt{\frac{1}{3}}  & \sqrt{\frac{1}{2}} },
\ee
and the low energy neutrino masses have this following form,
\be
D_k=y_{\nu}^2v_2^2 \rm{diag}(\frac{1}{a+b}, \frac{1}{a},\frac{1}{b-a} ).
\ee
Since in our model R parity is violated spontaneously, the neutral fermion  sector of our model will change significantly. Because of R parity violation the standard model neutrinos $\nu_i$ as well as the gauge singlet  Majorana neutrinos $N_i$ mix with the different neutralinos $\tilde{\lambda}^{0,3}$ and  $\tilde{H}_{u}^{0}$ \footnote{Being gauge singlet, $N_i$ does not mix with the gauginos $\tilde{\lambda}^{0,3}$, however mix with the higgsino $\tilde{H}_{u}^{0}$.}. In addition to this we also have neutrino-gauge  singlet higgsino mixings and the usual conventional standard model neutrino-gauge singlet  neutrino, MSSM gaugino-higgsino  and gauge singlet higgsino-higgsino mixings. In our model the neutral fermion basis is  $\psi=(\nu^{\prime},N^{\prime},{\Phi_1},{\Phi_2}, \tilde{\chi})$, where $\nu^{\prime}=(\nu_e,\nu_{\mu},\nu_{\tau})$, $N^{\prime}=(N_1,N_2,N_3)$, $\Phi_1=(\tilde{\xi},\tilde{\phi}_{S_1},\tilde{\phi}_{S_2},\tilde{\phi}_{S_3})$, $\Phi_2= (\tilde{\phi}_{T_1},\tilde{\phi}_{T_2},\tilde{\phi}_{T_3})$ and $\tilde{\chi}=(\tilde{\lambda}^0,\tilde{\lambda}^3,\tilde{H}^0_d,\tilde{H}^0_u)$. We denote the vacuum expectation values of the different Higgs and the sneutrino fields by $\langle \phi_{S_i}\rangle= u_i$, $\langle \phi_{T_i}\rangle= t_i$, $\langle \xi \rangle =s$, $\langle H^0_{d,u} \rangle =v_{1,2}$, $\langle \tilde{N}_i \rangle= w_i$ and  $\langle \tilde{\nu}_{L_i} \rangle=x_i$  respectively. The neutral fermion mass matrix is,
\be
L=-\frac{1}{2} \psi^T M \psi +h.c ,
\ee
where 
\be
M=\pmatrix{0 & A & 0 & 0 & B \cr A^T & C & D & 0 & P \cr 0 & D^T & K & 0 & 0 \cr 0 & 0 & 0 & F & 0 \cr B^T & P^T & 0 & 0 & G}.
\label{eq:mnufull}
\ee
In the above matrix $A$, $C$, $D$, $F$, $K$, $G$, $B$ and $P$ are these following matrices, 
\be
A= \pmatrix{y_{\nu}v_2 & 0 & 0 \cr 0 & 0 & y_{\nu}v_2 \cr 0 & y_{\nu}v_2 & 0 },
\ee
\be
C=2\pmatrix{ a+\frac{2b_1}{3} & -\frac{b_3}{3} & -\frac{b_2}{3} \cr  -\frac{b_3}{3} & \frac{2b_2}{3} & a-\frac{b_1}{3} \cr -\frac{b_2}{3} & a-\frac{b_1}{3} & \frac{2b_3}{3}},
\ee
\be
D=2\pmatrix{x_Aw_1 & \frac{2}{3}x_Bw_1 & -\frac{1}{3}w_3x_B & -\frac{1}{3}w_2x_B \cr x_Aw_3 & -\frac{1}{3}x_Bw_3 & \frac{2}{3}w_2x_B & -\frac{1}{3}w_1x_B \cr x_Aw_2 & -\frac{1}{3}x_Bw_2 & -\frac{1}{3}w_1x_B & \frac{2}{3}w_3x_B },
\ee
\be
F=2\pmatrix{\gamma+2\gamma^{\prime}t_1 & -\gamma^{\prime}t_3 & -\gamma^{\prime}t_2 \cr -\gamma^{\prime}t_3 & 2\gamma^{\prime}t_2 & \gamma-\gamma^{\prime}t_1 \cr -\gamma^{\prime}t_2 & \gamma-\gamma^{\prime}t_1 & 2\gamma^{\prime}t_3},
\ee
\be
K=2\pmatrix{ 3\alpha s  & \beta u_1  & \beta u_3 & \beta u_2 \cr \beta u_1 & \beta s+2 \beta^{\prime}u_1 & -\beta^{\prime}u_3  & -\beta^{\prime}u_2 \cr \beta u_3 & -\beta^{\prime}u_3 &  2\beta^{\prime} u_2 &  \beta s -\beta^{\prime}u_1 \cr \beta u_2  & -\beta^{\prime}u_2 & \beta s -\beta^{\prime}u_1 &  2\beta^{\prime} u_3 },
\ee
\be
G= \frac{1}{\sqrt{2}}\pmatrix{ \sqrt{2}M_1 & 0 & -{g_1v_1}& {g_1v_2} \cr
0& \sqrt{2}M_2 & {g_2 v_1}& -{g_2 v_2} \cr
-{g_1 v_1}& {g_2 v_1}& 0 & -\sqrt{2}\mu   \cr
{g_1 v_2} & -{g_2 v_2} & -\sqrt{2}\mu & 0  },
\label{eq:masnm}
\ee
\be
B=\frac{1}{\sqrt{2}}\pmatrix{ -g_1x_1 & g_2x_1 & 0 & \sqrt{2}y_{\nu}w_1 \cr -g_1x_2 & g_2x_2 & 0 & \sqrt{2}y_{\nu}w_3 \cr -g_1x_3 & g_2x_3 & 0 & \sqrt{2}y_{\nu}w_2},
\ee
and
\be
P=\pmatrix {0 & 0 & 0 & y_{\nu}x_1 \cr 0 & 0 & 0 & y_{\nu}x_3 \cr 0 & 0 & 0 & y_{\nu}x_2}.
\ee
Note that in the R parity conserving scenario the sneutrino vacuum expectation values  $x_i=0$ and $w_i=0$ and hence  the matrices $D$, $B$ and $P$ vanish. As a result the standard model light neutrinos $\nu_i$ as well as the gauge singlet neutrinos $N_i$ decouple from the the MSSM neutralino states  $\tilde{\chi}=(\tilde{\lambda}^{0,3}, \tilde{H}^0_{u,d})$ and as well as from the gauge singlet higgsino states  $\tilde{\phi}_T$, $\tilde{\phi}_S$, $\tilde{\chi}$,  as already have been mentioned previously. 

\section{Neutrino Mass and Tribimaximal Mixing}

In this section we discuss about the low energy neutrino phenomenology. We would like to stress that even with R-parity violation and enlargement of the neutrino sector as shown in Eq. (\ref{eq:mnufull}) the low energy neutrino mixing matrix will still  be the tribimaximal one,  provided the different sneutrino vacuum expectation values of $\tilde{N}_i$ and $\tilde{\nu}_i$ satisfies a particular equality relation, i.e  $w_1=w_2=w_3$ and $x_1=x_2=x_3$ along with the other necessary condition $u_1=u_2=u_3$. The mass matrix given in Eq. (\ref{eq:mnufull}) could be written as,
\be
M=\pmatrix{ 0 & M_D \cr M^T_D & M^{\prime}},
\ee
where $M_D$ is the  $ 3\times 14 $ block, $M^{\prime}$ is a $14 \times 14 $ matrix. Written in this way, the $M_D$ and $M^{\prime}$ are these following two matrices,
\be
M_D=\pmatrix {A & 0 & 0 & B},
\ee and 
\be
M^{\prime}=\pmatrix{C & D & 0 & P \cr D^T & K & 0 & 0 \cr 0 & 0 & F & 0 \cr P^T & 0 & 0 & G}.
\ee
The low energy neutrino mass matrix is,
\be 
M_{\nu} \sim M_D {M^{\prime}}^{-1} M^T_D.
\ee
Here we present the analytic expression of the low energy neutrino mass matrix. For the sake of simplicity we consider the couplings $x_A$, $x_B$, $\alpha$, $\beta$, $\beta^{\prime}$ of Eq. (\ref{eq:wsup})  to be the same. In addition to this we also consider the vacuum expectation values of $\phi_S$ and $\xi$ fields to be the same, i.e  $s=u$, however we have checked explicitly and numerically  that neither such a kind of VEV alignment between $s$ and $u$ nor a equality relation between the different couplings of the superpotential  is  a necessary criteria to get the tribimaximal mixing. With this above mentioned simplified assumption, the low energy neutrino mass matrix has this following form,
\be
M_{\nu}= \frac{2}{V} \pmatrix{8Q+3Z & 3Z-4Q & 3Z-4Q \cr 3Z-4Q & 3Z+5Q-\frac{24u^2Q}{w^2} & 3Z-Q+\frac{24u^2Q}{w^2} \cr 3Z-4Q & 3Z-Q+\frac{24u^2Q}{w^2} & 3Z+5Q-\frac{24u^2Q}{w^2}},
\label{eq:mnuapp}
\ee
where $Q$, $Z$ and $V$ are,
\be
Q=\mu u v_2^2y_{\nu}^2(\mu M_1M_2-(g_2^2M_1+g_1^2M_2)v_1v_2),
\ee
\be
Z=\beta(g_2^2M_1+g_1^2M_2)(8u^2-w^2)(\mu x+v_1wy_{\nu})^2,
\ee
\be
V=12\beta \mu(\mu M_1M_2-(g_2^2M_1+g_1^2M_2)v_1v_2)(8u^2-w^2).
\ee
Clearly $M_{\nu}(1,1)+M_{\nu}(1,2)=M_{\nu}(2,2)+M_{\nu}(2,3)$ and hence one expects the mixing matrix to be the tribimaximal one. Since $M_{\nu}$ is  a complex symmetric matrix, it will satisfy the diagonalizing relation  $U^TM_{\nu}U^*=D_k$, where $D_k=diag(m_1, m_2, m_3)$.  The low energy neutrino masses and mixing matrix which will come from the diagonalization of Eq. (\ref{eq:mnuapp}) are as follows, 
\be 
m_1=\frac{2uv_2^2y_{\nu}^2}{\beta(8u^2-w^2)}, \nn \\
m_3=-\frac{uv_2^2y_{\nu}^2}{\beta w^2}, \nn \\
m_2=\frac{3(g_2^2M_1+g_1^2M_2)(\mu x+v_1wy_{\nu})^2}{2\mu(\mu M_1M_2-(g_2^2M_1+g_1^2M_2)v_1v_2)}.
\ee
and the mixing matrix has this tribimaximal form,
\be
U=\pmatrix { \sqrt{\frac{2}{3}} & \frac{1}{\sqrt{3}} & 0 \cr -\frac{1}{\sqrt{6}} & \frac{1}{\sqrt{3}} & -\frac{1}{\sqrt{2}} \cr  -\frac{1}{\sqrt{6}} & \frac{1}{\sqrt{3}} & \frac{1}{\sqrt{2}}} .
\ee

\section{ Symmetry Breaking}

In this section we analyze the potential and the minimization  conditions. We show that in this model it is possible to get a higher value of $ \tilde{N} $ vacuum expectation value even for a smaller  vacuum expectation value of  the $\tilde{\nu}$ sneutrino field. The small value of $\langle \tilde{\nu} \rangle $ could be realized for small yukawa $y_{\nu}$. In our model the potential is,
\be
V=V_D+V_F+V_{soft}
\ee
where $V_D$, $V_F$  represent  the D term and F term contributions to the potential respectively and $V_{soft}$ comes from the soft supersymmetry breaking Lagrangian given in Eq. (\ref{eq:Lsoft}) and in Eq. (\ref{eq:softm}). The F term contribution is  $V_F=\sum_i|F_i|^2$, the index $i$ represents the different auxiliary fields of the theory. The neutral component of the potential which would be relevant for the analysis of symmetry breaking is $V_{neutral}=V^n_D+V^n_F+V^n_{soft}$ where the D term contribution to $V_{neutral}$ is,
\be
V^n_D=\frac{1}{8}(g_2^2+g_1^2)(|H^0_d|^2-|H^0_u|^2+\sum_i|\tilde{\nu}_i|^2)^2.
\ee
Note that the $A_4$ invariant K\"ahler potential of the gauge singlet matter chiral superfields $\hat{N}$ is, \footnote{$\hat{N}^{\dagger}\hat{N}= \hat{N}_1^{\dagger}\hat{N}_1+\hat{N}_2^{\dagger}\hat{N}_2+\hat{N}_3^{\dagger}\hat{N}_3$ is $A_4$ invariant, hence the K\"ahler potential is canonical.} 
\be
\mathcal K^N=  \int d^4\theta \hat{N}^{\dagger}\hat{N}.
\ee
Since $\hat{N}$ is singlet under the MSSM gauge group, hence it does not contributes to the $V_D$. The same argument  holds for the other  gauge singlets Higgs chiral superfield $\hat{\phi}_{T,S}$ and $\hat{\xi}$. 
The different auxiliary fields  $F_i$ which would be relevant in determining the neutral component of the potential are,  
\be
-F^*_{H^0_u}=y_{\nu}(\tilde{N}_1\tilde{\nu}_e+\tilde{N}_2\tilde{\nu}_{\tau}+\tilde{N}_3\tilde{\nu}_{\mu})-\mu H^0_d+...,
\ee
\be
-F^*_{H^0_d}=-\mu H^0_u+...,
\ee
\begin{eqnarray}
-F^*_{\nu_e}=y_{\nu}\tilde{N}_1H^0_{u}+..., \nn \\
-F^*_{\nu_{\mu}}=y_{\nu}\tilde{N}_3H^0_{u}+..., \nn \\
-F^*_{\nu_{\tau}}=y_{\nu}\tilde{N}_2H^0_{u}+...,
\end{eqnarray}
\be
-F^*_{\xi}=x_A(\tilde{N}_1\tilde{N}_1+2\tilde{N}_2\tilde{N}_3)+3\alpha \xi^2+\beta(\phi_{S_1}\phi_{S_1}+2\phi_{S_2}\phi_{S_3}),
\ee
\begin{eqnarray}
-F^*_{N_1}=2x_A\xi\tilde{N}_1+\frac{2x_B}{3}\big(2\phi_{S_1}\tilde{N}_1-\phi_{S_2}\tilde{N}_3-\phi_{S_3}\tilde{N}_2\big )+y_{\nu}\tilde{\nu}_eH_u^0,\\ \nn
-F^*_{N_2}=2x_A\xi\tilde{N}_3+\frac{2x_B}{3}\big(2\phi_{S_2}\tilde{N}_2-\phi_{S_1}\tilde{N}_3-\phi_{S_3}\tilde{N}_1\big )+y_{\nu}\tilde{\nu}_{\tau}H_u^0,\\ \nn
-F^*_{N_3}=2x_A\xi\tilde{N}_2+\frac{2x_B}{3}\big(2\phi_{S_3}\tilde{N}_3-\phi_{S_1}\tilde{N}_2-\phi_{S_2}\tilde{N}_1\big )+y_{\nu}\tilde{\nu}_{\mu}H_u^0,\\ \nn
\end{eqnarray}
\begin{eqnarray}
-F^*_{S_1}=\frac{2x_B}{3}\big( \tilde{N}_1\tilde{N}_1-\tilde{N}_2\tilde{N}_3)+ 2\beta \phi_{S_1}\xi+ 2\beta^{\prime}(\phi^2_{S_1}-\phi_{S_2}\phi_{S_3}),   \\ \nn
-F^*_{S_2}=\frac{2x_B}{3}\big( \tilde{N}_2\tilde{N}_2-\tilde{N}_1\tilde{N}_3)+ 2\beta \phi_{S_3}\xi+ 2\beta^{\prime}(\phi^2_{S_2}-\phi_{S_1}\phi_{S_3}),   \\ \nn
-F^*_{S_3}=\frac{2x_B}{3}\big( \tilde{N}_3\tilde{N}_3-\tilde{N}_1\tilde{N}_2)+ 2\beta \phi_{S_2}\xi+ 2\beta^{\prime}(\phi^2_{S_3}-\phi_{S_1}\phi_{S_2}),   
\end{eqnarray}
and 
\begin{eqnarray}
-F^*_{T_1}=2\gamma^{\prime}(\phi^2_{T_1}-\phi_{T_2}\phi_{T_3})+2\gamma \phi_{T_1}+...,\\ \nn
-F^*_{T_2}=2\gamma^{\prime}(\phi^2_{T_2}-\phi_{T_1}\phi_{T_3})+2\gamma \phi_{T_3}+..., \\ \nn
-F^*_{T_3}=2\gamma^{\prime}(\phi^2_{T_3}-\phi_{T_1}\phi_{T_2})+2\gamma \phi_{T_2}+...
\end{eqnarray}
With these auxiliary fields of the superfields $\hat{H}_{u,d}$, $\hat{\nu}_i$, $\hat{\xi}$, $\hat{N}$, $\hat{\phi}_S$ and $\hat{\phi}_T$ the F term contribution  to the neutral component of the potential will be,
\be
V^n_F=V^{F,n}_{H^0_u,H^0_d,\tilde{\nu}_i}+V^{F,n}_{\xi} + V^{F,n}_{N}+V^{F,n}_{S}+V^{F,n}_{T},
\label{eq:vneu}
\ee
where 
\be
\langle V^{F,n}_{H^0_u,H^0_d,\tilde{\nu}_i} \rangle &=& \mu^2 v_1^2+ \mu^2 v_2^2+ y_{\nu}^2 (w_1x_1+w_2x_3+w_3x_2)^2\nn \\
&&  -2\mu y_{\nu}v_1 (w_1x_1+w_2x_3+w_3x_2) +y_{\nu}^2v_2^2 (w_1^2+w_2^2+w_3^2),
\ee
\be
\langle V^{F,n}_{\xi} \rangle = |x_A(w_1^2+2w_2w_3)+3\alpha s^2+\beta(u_1^2+2u_2u_3)|^2 ,
\ee
\be
\langle V^{F,n}_{N} \rangle &=& |2x_Asw_1+\frac{2x_B}{3}(2u_1w_1-u_2w_3-u_3w_2)+y_{\nu}x_1v_2|^2 \\ \nn && +|2x_Asw_3+\frac{2x_B}{3}(2u_2w_2-u_1w_3-u_3w_1)+y_{\nu}x_3v_2|^2 \\ \nn && +|2x_Asw_2+\frac{2x_B}{3}(2u_3w_3-u_1w_2-u_2w_1)+y_{\nu}x_2v_2 |^2, 
\ee
\be
\langle V^{F,n}_{S} \rangle &=& |\frac{2x_B}{3}(w_1^2-w_2w_3)+2\beta u_1s+2\beta^{\prime}(u_1^2-u_2u_3)|^2 \nn \\ && + |\frac{2x_B}{3}(w_2^2-w_1w_3)+2\beta u_3s+2\beta^{\prime}(u_2^2-u_1u_3)|^2 \nn \\ && +|\frac{2x_B}{3}(w_3^2-w_1w_2)+2\beta u_2s+2\beta^{\prime}(u_3^2-u_1u_2)|^2,
\ee
and
\be
\langle V^{F,n}_{T} \rangle &=& | 2\gamma^{\prime}(t_1^2- t_2t_3)+ 2\gamma t_1 |^2+  | 2\gamma^{\prime}(t_2^2- t_1t_3)+ 2\gamma t_3 |^2 \nn \\ && +|2\gamma^{\prime}(t_3^2- t_1t_2)+ 2\gamma t_2 |^2.
\ee
The soft supersymmetry breaking Lagrangian for this model has been given in Eq. (\ref{eq:Lsoft}) and Eq. (\ref{eq:softm}) in the appendix and  below we write the soft supersymmetry breaking contribution to $\langle V_{neutral} \rangle $,  
\be
\langle V^n_{soft} \rangle &=& 2\tilde{y}_{\nu}(w_1x_1+w_2x_3+w_3x_2)v_2+2\tilde{x}_As(w_1^2+2w_2w_3) \nn \\ && +\frac{2\tilde{x}_B}{3}(2u_1w_1^2+2u_2w_2^2+2u_3w_3^2-2u_1w_2w_3-2u_2w_1w_3-2u_3w_1w_2)\\ \nn  && +2\tilde{\alpha}s^3 
+2\tilde{\beta}s(u_1^2+2u_2u_3)+2\tilde{\gamma}(t_1^2+2t_2t_3)+\frac{4\tilde{\beta}^{\prime}}{3}(u_1^3+u_2^3+u_3^3-3u_1u_2u_3)\nn \\ && +\frac{4\tilde{\gamma}^{\prime}}{3}(t_1^3+t_2^3+t_3^3-3t_1t_2t_3)+r_{\xi}s^2+r_T(t_1^2+t_2^2+t_3^2)+r_S(u_1^2+u_2^2+u_3^2)\nn \\ && +r_N(w_1^2+w_2^2+w_3^2)+ m^2_{H_u}v_2^2+ m^2_{H_d}v_1^2-2bv_1v_2+m^2_{\tilde{L}}(x_1^2+x_2^2+x_3^2).
\ee
To simplify the analysis we assume all the vacuum expectation values and the couplings as real and also assume $u_1=u_2=u_3=u$, $w_1=w_2=w_3=w$, $x_1=x_2=x_3=x$ and $t_1=t, t_2=t_3=0$. Minimizing $\langle V_{neutral} \rangle$ given in Eq. (\ref{eq:vneu}) w.r.t  the different vacuum expectation values $t$, $w$, $x$, $s$, $u$ and  $v_{1,2}$ we get these following equations respectively,
\be
8 {\gamma^{\prime}}^2t^2+ 4\gamma^2+ 12\gamma^{\prime}\gamma t+ r_T+2\tilde{\gamma}^{\prime}t+2\tilde{\gamma}=0;  t\neq 0,
\label{eq:mint}
\ee
\be
36x_A^2w^3+36x_A\alpha s^2w+36x_A\beta w u^2+18y_{\nu}^2x^2w-6y_{\nu}\mu v_1x+6y_{\nu}^2v_2^2w \nn \\ 12x_Ay_{\nu}v_2sx+24x_A^2s^2w+6\tilde{y}_{\nu}v_2x+12\tilde{x}_Asw+6r_Nw=0,
\label{eq:minw}
\ee
\be
18y_{\nu}^2w^2x-6y_{\nu} \mu v_1w+6 y_{\nu}^2v_2^2x+6 \tilde{y}_{\nu}v_2w+6 m^2_{\tilde{L}}x \nn \\  
+12x_Ay_{\nu}v_2sw+\frac{(g_1^2+g_2^2)}{2}3x(v_1^2-v^2_2+3x^2)=0,
\label{eq:minx}
\ee
\be
36\alpha^2s^3+36x_A\alpha s w^2+ 36 \beta \alpha s u^2+ 24 x_A^2 s w^2+ 24 \beta^2 u^2 s \\ \nn +12x_Ay_{\nu}v_2wx+ 6 \tilde{x}_Aw^2+ 6\tilde{\alpha}s^2+6\tilde{\beta}u^2+2r_{\xi}s=0,
\label{eq:mins}
\ee
\be
36 \beta^2 u^3+36 \beta \alpha s^2u+36 x_A\beta w^2u+24\beta^2u s^2+12 \tilde{\beta}su+6r_Su=0,
\label{eq:minu}
\ee
\be
2\mu^2v_1+2m^2_{H_d}v_1-2bv_2-6y_{\nu} \mu w x+\frac{v_1}{2}(g_1^2+g_2^2)(v_1^2-v_2^2+3x^2)=0,
\label{eq:minv1}
\ee
\be
2\mu^2v_2+2m^2_{H_u}v_2-2bv_1-\frac{v_2}{2}(g_1^2+g_2^2)(v_1^2-v_2^2+3x^2) \nn \\ 
+12x_Ay_{\nu}swx+6y_{\nu}^2w^2v_2+6y_{\nu}^2x^2v_2+6\tilde{y}_{\nu}wx=0.
\label{eq:minv2}
\ee
As evident from Eq. (\ref {eq:minx}), the vacuum expectation value $x \to 0$ could be naturally realized in the limit  $y_{\nu} \to 0$. In the  limit of small yukawa i.e $y_{\nu} \to 0$  and $x \to 0$  from Eq. (\ref{eq:minw})  one will get the following relation between the  different vacuum expectation values  of the  sneutrino $\tilde{N}$ and the Higgs  $\phi_S$ and $\xi$ fields,
\be
w^2=-\frac{6x_A\alpha s^2+6x_A\beta u^2+4 x_A^2s^2+2\tilde{x}_A s+r_N}{6x_A^2}
\label{eq:wsq}
\ee
Using Eq. (\ref{eq:minw}) and  Eq. (\ref{eq:minu}), the vacuum expectation value $s$ of the $\xi$ field would come as,
\be
s=\frac{-\tilde{B}\pm \sqrt{\tilde{B}^2-4\tilde{A}\tilde{C}}}{2\tilde{A}}
\ee
where 
$\tilde{A}=4x_A\beta(x_A-\beta)$, $\tilde{B}=2(\beta \tilde{x}_A-\tilde{\beta}x_A)$ and $\tilde{C}=(\beta r_N-r_Sx_A)$.
It is clearly evident from Eq. (\ref{eq:wsq}) that  the vacuum expectation value $w$ of the sneutrino field $\tilde{N}$ is related with the vacuum expectation value of the different standard model singlet R-Parity even Higgs fields $\xi$ and $\phi_S$, as well as it depends on the soft supersymmetry breaking coupling $r_N$ and even in the $\langle \tilde{\nu} \rangle =x \to 0$ limit it is possible to get a nonzero and higher value of the sneutrino vacuum expectation value $\langle \tilde{N} \rangle$.

\section{Conclusion}

In this work we have explored the possibility of spontaneous R parity violation in the context of a specific flavor model. The flavor symmetry group is $A_4$ and in addition we have implemented another discrete symmetry group $Z_3$. The superpotential given in Eq. (\ref{eq:wsup}) conserves R-parity. However R-parity would be broken spontaneously when the different sneutrino fields which are odd under R parity will get the vacuum expectation values. The $A_4$ flavor symmetry will be broken by the vacuum expectation values of the $A_4$ triplet fields. Because of the R parity violation we have mixing between the standard model neutrinos,  MSSM higgsinos and gauginos, as well as mixing between the gauge singlet  neutrinos and MSSM higgsinos. In  our model the Higgs and  higgsino sector is enlarged  because of the presence of the gauge singlet Higgs chiral superfield $\hat{\phi}_S$, $\hat{\phi}_T$ and $\hat{\xi}$. Hence, in addition to the higgsino-higgsino, gaugino-higgsino and the conventional Dirac type neutrino-neutrino mixings we also have mixings between the neutrino and these  gauge  singlet higgsinos. As a result in our model the neutral fermion mass matrix is a $17 \times 17 $ matrix. However we show that the low energy neutrino mass matrix which will be generated once the different heavier neutral fermionic states are integrated out, can still have a specific form leading to the tribimaximal mixing matrix, provided  few  constraints between the different sneutrino and Higgs vacuum expectation values are satisfied.

We have also explored the potential minimization in detail. One can relate the spontaneous R parity violation  with some higher gauge symmetry breaking and in these kind of models \cite{pavel} one realizes a higher value of one of the the R parity violating sneutrino vacuum expectation value. On the other hand if one sticks to the basic MSSM gauge group and also the MSSM particle content and explores the R parity violation, one would eventually get into the trouble  of Majoron\cite{spontaneous}. However introducing one MSSM gauge group singlet/triplet matter chiral superfield one can avoid the problem of Majoron because of the explicit breaking of lepton number although breaking R parity spontaneously\cite{Choubey:2009wp}. In this kind of models  the different sneutrino vacuum expectation values  share a proportionality relation and  the smallness of the neutrino mass forces the sneutrino vacuum expectation values to be small. In this present model although we stick to the basic MSSM gauge group, however it is possible to realize a higher value of the sneutrino VEV $\langle \tilde{N} \rangle$ even in the $\langle \tilde{\nu} \rangle \to 0$ limit. We have analyzed the potential and have shown this particular feature explicitly. For the sake of simplicity we have assumed the different sneutrino vacuum expectation values $w_{1,2,3}= w$ and $x_{1,2,3}= x$ as well as the vacuum expectation values of $\phi_S$ fields $u_{1,2,3} = u$. Note that for this assumption one would also obtain the desired tribimaximal mixing in the low energy neutrino sector, as already been mentioned in section 3 and in the earlier paragraph.  We have shown that the sneutrino vacuum expectation value $\langle \tilde{N} \rangle$ is related with the other Higgs vacuum expectation values  $\langle \tilde{\phi}_s \rangle$ and  $\langle \tilde{\xi} \rangle$ as well as it depends on the soft supersymmetry breaking parameter $r_N$ given in Eq. (\ref{eq:softm}). Hence for $r_N \sim \rm{TEV} $  one would also expect  $\langle \tilde{N} \rangle$ in the TeV scale.

\vglue 0.8cm
\noindent
{\Large{\bf Acknowledgments}}\vglue 0.3cm
\noindent
The author would like to thank Sandhya Choubey,  Amitava Raychaudhuri, Priyotosh Bandyopadhyay  and Srubabati Goswami for useful discussions, comments   and support.
This work has been supported by the Neutrino Project 
under the XI Plan of Harish-Chandra Research Institute (HRI).  
The authors acknowledge the HRI 
cluster facilities for computation.

\vglue 1.0cm

\begin{center}
{\LARGE{\bf \underline {Appendix}}}
\end{center}

\begin{appendix}
\section{Soft Supersymmetry Breaking}
Here we write the soft supersymmetry breaking Lagrangian of our model. Following supergravity mediated supersymmetry breaking ansatz \cite{msugra} \footnote{ We consider the canonical  K\"ahler potential and the gauge kinetic function to be $f_{AB}=\delta_{AB}$} the soft supersymmetry breaking lagrangian of this model would be 
\be
\mathcal L^{\rm{soft}} = \mathcal L_{\rm{1}}^{\rm{soft}}+ \mathcal L^{(\phi_{T},\phi_{S},\xi,N)}_{soft}
\ee
where 
\be
-\mathcal L_{\rm{1}}^{\rm{soft}} &=&
(m_{\tilde{Q}}^2)^{ij} {\tilde Q^{\dagger}_i} \tilde{Q_j}
+(m_{\tilde u^c}^{2})^{ij}
{\tilde u^{c^*}_i} \tilde u^c_j
+(m_{\tilde d^c}^2)^{ij}{\tilde d^{c^*}_i}\tilde d^c_j
+(m_{\tilde{L}}^2)^{ij} {\tilde L^{\dagger}_i}\tilde{L_j} \nonumber \\
&+&(m_{\tilde e^c}^2)^{ij}{\tilde e^{c^*}_i}\tilde e^c_j 
+ m_{H_d}^2 {H^{\dagger}_d} H_d + m_{H_u}^2 {H^{\dagger}_u} H_u +(bH_uH_d+ \rm{H.c.}) \nn \\
&+&  \left[
-A_u^{ij} H_u\tilde Q_i \tilde u_j^c +
A_d^{ij} H_d \tilde Q_i \tilde d_j^c+ \text{H.c} \right ] \nonumber \\
&+& \left [\frac{\tilde{y}_1}{\Lambda} H_d \langle \phi_T \rangle \tilde{L}\tilde{e}^c  +\frac{\tilde{y}_2}{\Lambda} H_d  (\langle \phi_T \rangle \tilde{L})^{\prime \prime}\tilde{\mu}^c + \frac{\tilde{y}_3}{\Lambda} H_d (\langle \phi_T \rangle \tilde{L})^{\prime}\tilde{\tau}^c +
 \text{H.c.}  \right] 
\nonumber \\
 &+&\frac{1}{2} \left(M_3 \tilde{g} \tilde{g}
+ M_2 \tilde{\lambda}^i \tilde{\lambda}^i + M_1 \tilde{\lambda}^0 
\tilde{\lambda}^0 + \text{H.c.} \right).
\label{eq:Lsoft}
\ee
and
\be
-\mathcal L^{(\phi_{T},\phi_{S},\xi,N)}_{soft}&=& r_{\xi}\xi^{\dagger}\xi+ r_T \phi_T^{\dagger}\phi_T+ r_S \phi_S^{\dagger}\phi_S+r_N \tilde{N}^{\dagger}\tilde{N} \nn \\ && +\big ( \tilde{y}_{\nu}\tilde{N}\tilde{L}{H}_u+\tilde{x}_A{\xi}\tilde{N}\tilde{N}+\tilde{x}_B{\phi}_S\tilde{N}\tilde{N}+\tilde{\alpha}{\xi}{\xi}{\xi} \nn \\ &&  +\tilde{\beta}^{\prime}{\phi}_S{\phi}_S{\phi}_S+\tilde{\beta}{\phi}_S{\phi}_S{\xi} +\tilde{\gamma}{\phi}_T{\phi}_T +\tilde{\gamma}^{\prime}{\phi}_T{\phi}_T{\phi}_T+h.c \big )
\label{eq:softm}
\ee
Note that the trilinear soft supersymmetry breaking terms involving the different slepton fields would be generated from the $ \frac{\tilde{y}_1}{\Lambda} H_d (\phi_T\tilde{L})\tilde{e}^c$, $ \frac{\tilde{y}_2}{\Lambda} H_d(\phi_T\tilde{L})^{\prime \prime}\tilde{\mu}^c$ and $\frac{\tilde{y}_3}{\Lambda}H_d(\phi_T\tilde{L})^{\prime}\tilde{\tau}^c $ operators once the $A_4$ triplet  Higgs $\phi_T$ gets the vacuum expectation value, thereby breaking $A_4$ spontaneously. Since we adopt a supergravity mediated supersymmetry breaking mechanism, $\tilde{y}_{\nu}= y_{\nu}a$ and similar kind of relation would hold for other trilinear and bilinear couplings as well. The K\"ahler potential involving the gauge singlet superfields $\hat{\phi}_{S,T}$,  $\hat{\xi}$ and $\hat{N}$ is,
\be
\mathcal K^S=\int d^4\theta  (\hat{\phi}_S^{\dagger}\hat{\phi}_S+\hat{\phi}_T^{\dagger}\hat{\phi}_T+\hat{\xi}^{\dagger}\hat{\xi}+\hat{N}^{\dagger}\hat{N})
\label{eq:kah1}
\ee
The K\"ahler potential for the other superfields which transform nontrivially under standard model gauge group will have this form,
\be
\mathcal K=\int d^4\theta (\hat{\phi}_i^{\dagger}e^{2gV}\hat{\phi}_i)
\label{eq:kah2}
\ee
where $\hat{\phi}_i$ is any MSSM superfield and $V$ is the vector superfield.
\end{appendix}

\end{document}